\documentclass[authoryear,hyperref,preprint,1p]{elsarticle}
\makeatletter
\def\ps@pprintTitle{%
     \let\@oddhead\@empty
     \let\@evenhead\@empty
     \def\@oddfoot{}%
     \let\@evenfoot\@oddfoot}
\makeatother
\RequirePackage{color}
\RequirePackage{url}
\RequirePackage{ifthen}
\usepackage{lineno}
\usepackage{amsmath}
\usepackage{subfigure}
\usepackage{amssymb}
\RequirePackage{multirow}
\RequirePackage{hangcaption}
\newtheorem{Thm}{Theorem}[section]   

\newdefinition{Rem}[Thm]{Remark}

\newcommand{\rfia}[1]{\makebox[\parindent][l]{%
                     \makebox[0em][r]{\rm(}\sf#1\rm)}}
\newcounter{ABCcB}
\newcommand{\theABCcC}{\alph{ABCcB}}

\newenvironment{ABC}{\begin{list}{
  \rfia{\theABCcC}}{\usecounter{ABCcB} \topsep 0ex \partopsep 0ex \itemsep0ex
  \parsep=\parskip \leftmargin 0em \rightmargin 0em \itemindent=\parindent
  \listparindent=\parindent  \labelsep 0.5em \labelwidth 0.5em }}{\end{list}}

\newcommand{\href}[2]{{#2}}
\newcommand{\hreft}[1]{{\href{#1}{\tt\url{#1}}}}

\newcommand{\Ew}{\mathop{\rm {{}E{}}}\nolimits} 
\newcommand{\Var}{\mathop{\rm Var}\nolimits}    
\newcommand{\Cov}{\mathop{\rm Cov}\nolimits}    
\newcommand{\ve}{\varepsilon}

\newcommand{\SSs}{\scriptscriptstyle}
\newcommand{\Ts}{\textstyle}

\newcommand{\ssr}{\rm\scriptscriptstyle}

\newcommand{\tr}{\mathop{\rm{} tr{}}}

\newcommand{\R}{\mathbb R}

\newcommand{\N}{\mathbb N}
\newcommand{\EM} {{\mathbb I}}

\newcommand{\iid}{\mathrel{\stackrel{\ssr i.i.d.}{\sim}}}

\newcommand{\Tfrac}[2]{\textstyle\frac{#1}{#2}}

\newlength{\SyW}

\begin{document}
\begin{frontmatter}
\title
{Optimally Robust Kalman Filtering at Work: AO-, IO-, and
 Simultaneously IO- and AO- Robust Filters}
\author{Peter Ruckdeschel}
\ead{Peter.Ruckdeschel@itwm.fraunhofer.de}
\date{\today}
\address{Fraunhofer ITWM, Abt.\ Finanzmathematik,
         Fraunhofer-Platz 1, 67663 Kaiserslautern, Germany\\
     and TU Kaiserslautern, AG Statistik, FB.\ Mathematik,
         P.O.Box 3049, 67653 Kaiserslautern, Germany
}
\begin{abstract}
We take up optimality results for robust Kalman filtering
from \cite{Ru:01,Ru:10} where robustness is understood in
a distributional sense, i.e.; we enlarge the distribution
assumptions made in the ideal model by suitable neighborhoods,
allowing for outliers which in our context may be
system-endogenous/propagating or -exogenous/non-propagating,
inducing the somewhat conflicting goals of tracking and
attenuation. Correspondingly, the cited references provide
optimally-robust procedures to deal with each type of
outliers separately, but in case of IO-robustness does not
say much about the implementation. We discuss this in more
detail in this paper. Most importantly, we define a hybrid
filter combining AO- and IO-optimal ones, which is able to
treat both types of outliers simultaneously, albeit with a
certain delay.
We check our filters at a reference state space model, and
compare the results with those obtained by the ACM filter
\cite{Ma:Ma:77}, \cite{Ma:79} and non-parametric,
repeated-median based filters \cite{F:B:G:06}, \cite{F:E:G:07}.
\end{abstract}
\end{frontmatter}
%
%
\section{Introduction}
Robustness is an ``old'' problem in Kalman filtering, with first
(non-verified) hits on a quick search for ``robust Kalman filter'' on
\hreft{scholar.google.com} as early as 1962, i.e.; even before the
seminal \citet{Hu:64} paper, often referred to as birthday of Robust
Statistics.

The amount of literature on this topic is huge, and we do not attempt to
give a comprehensive account here. Instead, we refer the reader to the
excellent surveys given in \citet{Er:Li:78}, \citet{Ka:Po:85}, \citet{St:Du:87},
\citet{Sch:Mi:94}, \citet{Kue:01}, and to some extent \citet[Sect.~1.5]{Ru:01}.

The mere notion of robustness in the context of filtering is non-standard,
most frequently it is used to describe stability of the procedure w.r.t.\
certain variations of the ``input parameters''. The choice of
the ``input parameters'' to look at varies from notion to notion.
In this paper we are concerned with (distributional) {\bf minimax robustness}; i.e.;
we allow for deviations from the ideal distributional model assumptions,
defining suitable neighborhoods about this ideal model.
For these neighborhoods, we consider procedures minimizing the maximal predictive
inaccuracy on these neighborhoods, measured in terms of mean squared error
(MSE).

These minimax filters derived in \citet{Ru:10} come as closed-form saddle-points
consisting of an optimally-robust procedure and a corresponding least favorable
outlier situation. Their optimality holds in a surprisingly general setup of
state space models, which is not limited to a Euclidean or time-discrete framework.

The important condition is that MSE makes sense for the range of the states, so
these results cover general Hidden Markov Models for arbitrary observation space,
dynamic (generalized) linear models as discussed in \citet{W:H:M:85} and
\citet{We:Ha:89}, as well as continuous time settings, as those used in
applications of Mathematical Finance, compare, e.g.\ \citet{N:M:M:00} and
\citet{Si:02} for respective model formulations.
For the application, we need to linearize the corresponding functions transition
and observation functions suitably, e.g.\ to give the \emph{(continuous-discrete)
Extended Kalman Filter (EKF)}. We even cover applications such as
optimal portfolio selection, with corresponding controls $U_t$ (i.e.; buying and
selling operations) and, in principle, indirectly observed random fields.

In this paper though, to clarify ideas we limit ourselves to the linear,
Euclidean, time discrete state space model (SSM), and in the ideal model
assume normality. More details on these models can be found in
 many textbooks, cf.\ e.g.\ \citet{A:M:79}, \citet{Ha:91}, \citet{Ham:93},
and \citet{D:K:01}.

\section{Setup}\label{GenSetup}
\subsection{Ideal model} \label{idmodel}

Let us start with some definitions and assumptions.
%
Our SSM consists in an  unobservable $p$-dimensional
state $X_t$ evolving according to a possibly time-inhomo\-geneous
vector autoregressive model of order $1$  (VAR(1)) with
innovations $v_t$ and transition matrices $F_t$, i.e.,
\begin{equation} \label{VAR1}
X_t=F_t  X_{t-1}+ v_t
\end{equation}
We only observe a $q$-dimensional linear transformation
$Y_t$ of $X_t$ involving an additional observation error $\ve_t$,
\begin{equation}
Y_t =Z_t  X_t+ \ve_t \label{linobs}
\end{equation}
In the ideal model we work in a Gaussian context, that is we assume
\begin{align}
&  v_t \stackrel{\rm \SSs indep.}{\sim} {\cal N}_p(0,Q_t), \qquad
\ve_t \stackrel{\rm \SSs indep.}{\sim} {\cal N}_q(0,V_t),  \qquad
 X_0  \stackrel{\phantom{\rm \SSs indep.}}{\sim}  {\cal N}_p(a_0,Q_0),
 \label{normStart}\\
&{\mbox{$ \{X_0, v_s, \ve_t,  \;s,t\in\N\}$ stochastically independent}}
\end{align}
For this paper, we assume the hyper--parameters $F_t,Z_t,Q_t,V_t,a_0$ to be known.
%

\subsection{Deviations from the ideal model} \label{devidmod}
As announced, these ideal model assumptions for robustness considerations
are extended by allowing (small) deviations, most prominently generated
by outliers.
In our notation, suffix ``${\Ts \rm id}$'' indicates the {\it id}eal setting,
``${\Ts \rm di}$'' the {\it di}storting (contaminating) situation, ``${\Ts \rm re}$''
the {\it re}alistic, contaminated situation.

\paragraph{AO's and IO's}
We follow the terminology of \citet{F:72}, who distinguishes
{\em innovation outliers} (or IO's) and {\em additive outliers} (or AO's).
Historically, AO's and IO's denote gross errors affecting the observation
errors and the innovations, respectively. For consistency with literature,
we stick to this distinction, but rather use these terms in a wider sense:
\textit{IO}'s stand for general endogenous outliers entering the state
equation, hence with propagated distortion,  also covering level shifts
or linear trends which would not be included in the original definition.
Similarly wide-sense \textit{AO}'s denote general exogenous outliers
which do not propagate, like substitutive outliers or \textit{SO}'s as
defined in equations~\eqref{YSO}--\eqref{U-SO}.

\paragraph{Different and competing goals induced by AO's and IO's}

Due to their different nature, as a rule, a different reaction in
the presence of IO's and AO's is required. As AO's are exogenous, we would
like to ignore them as far as possible, damping their effect,
while when there are IO's, something has happened in the system, so
 the usual goal will be to  detect these  structural changes as fast as
 possible.

 A situation where both AO's and IO's may occur is more difficult: We
 are faced with an identification problem, as we cannot distinguish
 IO from AO type immediately after a suspicious observation; hence
 a simultaneous treatment is only possible with a certain
 delay---see section~\ref{IOAO'sec}.

 Another task, not pursued further in this paper, consists in
 recovering the situation without structural changes in an
 off-line situation. An example is spectral analysis of inter-individual
 heart frequency spectra, which requires a cleaning from both (wide-sense)
 IO's and AO's; after this cleaning the powerful instruments
 of spectral analysis will be available;  cf. \citet{Spa:08}.

\section{Kalman filter and robust alternatives} \label{rLSsec}
%
\subsection{Classical Method: Kalman--Filter} \label{classKalmss}

\paragraph{Filter Problem}
The most important problem in SSM formulation is the reconstruction of
the unobservable states $X_t$ by means of the observations $Y_t$.
For abbreviation let us denote
\begin{equation}
Y_{1:t}=(Y_1,\ldots,Y_t), \quad Y_{1:0}:=\emptyset
\end{equation}
Using MSE risk, the optimal reconstruction is the solution to
\begin{equation}
\Ew \big| X_t-f_t\big|^2 = \min\nolimits_{f_t},\qquad
  f_t \mbox{ measurable w.r.t.\ } \sigma(Y_{1:s})  \label{MSEpb}
\end{equation}
We focus on filtering ($s=t$) in this paper, while $s<t$ makes for a
prediction, and $s>t$ for a smoothing problem.
{\paragraph{Kalman--Filter}
The general solution to \eqref{MSEpb}, the corresponding
conditional expectation  $\Ew[ X_t|Y_{1:s}]$ in general is rather
expensive to compute. Hence as in  the Gauss-Markov setting,
restriction to linear filters is a common way out.
In this context, \citet{Kal:60} introduced a recursive scheme to
compute this optimal linear filter reproduced here for
later reference:
\begin{align}
\!\!\!\!\!\!\mbox{Initialization: }&\!\!\!\!&
 X_{0|0} &= a_0,\qquad &\Sigma_{0|0}&=Q_0 \label{bet1}
 \\
\!\!\!\!\!\!\mbox{Prediction: }&\!\!\!\!
& X_{t|t-1}&= F_t  X_{t-1|t-1} , \qquad &\Sigma_{t|t-1}
&=F_t\Sigma_{t-1|t-1}F_t^\tau + Q_t\label{bet2}
\\
\!\!\!\!\!\!\mbox{Correction: }
&\!\!\!\!& X_{t|t}&=  X_{t|t-1} + M^0_t \Delta Y_t, \label{bet3}
\!\!\!\quad&  \Delta Y_t &= Y_t-Z_t x_{t|t-1},\nonumber\\
&&M^0_t&=\Sigma_{t|t-1}Z_t^\tau \Delta_t^{-1},
\nonumber
\!\!\!\quad&  \Sigma_{t|t} &= (\EM_p-M^0_tZ_t)\Sigma_{t|t-1},\nonumber\\
&&\Delta_t&=Z_t \Sigma_{t|t-1}Z_t^\tau + V_t
\end{align}
where $\Sigma_{t|t}=\Cov(X_t-X_{t|t})$, $\Sigma_{t|t-1}=\Cov(X_t-X_{t|t-1})$,
and $M^0_t$ is the so-called \textit{Kalman gain}.

The Kalman filter has a clear-cut structure with an initialization, a
prediction, and a correction step. Evaluation and interpretation is
easy, as all steps are linear.
The strict recursivity / Markovian structure of the state equation
allows one to concentrate all information from the past useful for
the future in  $X_{t|t-1}$.\\
This linearity is also the reason for its non-robustness, as
observations $y$ enter unbounded into the correction step. A good
robustification has to be bounded in the observations, otherwise
preserving the advantages of the Kalman filer as
far as possible.
\subsection{The rLS as optimally robust filter} \label{rLSsec}
The idea of the procedures we discuss in this paper are based on
{{\it r}obustifying {\it r}ecursive  {\it L}east {\it S}quares: rLS}.
Let us begin with (wide-sense) AO's.

With only AO's, there is no need for robustification in the initialization
and prediction step, as no (new) observations enter. For the correction step,
we use the following robustification, compare \citet{Ru:00}:
Instead of $M^0\Delta Y$, we use a Huberization of this correction
\begin{equation} \label{HbDef}
H_b(M^0\Delta Y) =  M^0\Delta Y \min\{1, b/\big|M^0\Delta Y\big|\}
\end{equation}
for some suitably chosen clipping height $b$. While this is a bounded
substitute for the correction step in the classical Kalman filter, it still
remains reasonably simple, is non iterative and hence especially useful
for online-purposes.

 However it should be noted that, departing from the Kalman filter and
at the same time insisting on strict recursivity, we possibly exclude
``better'' non-re\-cur\-sive procedures. These procedures on the other
hand would be much more expensive to compute.

\begin{Rem}
  $\big|\,\cdot\,\big|$ in expression $\big|M^0\Delta Y\big|$ denotes
  the Euclidean norm of $\R^q$; instead, however you could also use
  other norms like Mahalanobis-type ones. The choice of a quadratic-form-type
  norm for $H_b$ does not affect the optimality statements of
  Theorem~\ref{ThmSO} below, provided that the same norm be used in the MSE.
\end{Rem}

\paragraph{Choice of the clipping height $b$}
For the choice of  $b$, we have two proposals. Both are based on the
simplifying assumption that $\Ew_{\rm\SSs id}  [\Delta X | \Delta Y]$
is linear, which in fact turns out to  only be approximately correct.
The first one chooses $b=b(\delta)$ according to an Anscombe criterion,
    \begin{equation}
     \Ew_{\rm\SSs id} \big|\Delta  X-H_b(M^0 \Delta Y)\big|^2
         \stackrel{!}{=}(1+\delta)
          \Ew_{\rm\SSs id} \big|\Delta  X-M^0 \Delta Y\big|^2
         \label{deltakrit}
    \end{equation}
where $\delta$ may be interpreted as ``insurance premium''
to be paid in terms of efficiency.

The second criterion uses the radius $r\in[0,1]$  of the
neighborhood ${\cal U}^{\rm\SSs SO}(r)$ (defined in \eqref{U-SO}) and
determines $b=b(r)$ such that
    \begin{equation}
     (1-r) \Ew_{\rm\SSs id} (|M^0 \Delta Y |-b)_+ \stackrel{!}{=} r b
         \label{deltakrit2}
    \end{equation}
This will produce the minimax-MSE procedure for ${\cal U}^{\rm\SSs SO}(r)$
according to Theorem~\ref{ThmSO} below.

If $r$ is unknown, which is almost always the case, we translate an idea
worked out in \citet{RKR08} into our setting: Assume we know that
$r\in[r_l,r_u]$, $0\leq r_l<r_u\leq 1$. Then we define a
\textit{least favorable radius} $r_0$ such that the
procedure ${\rm rLS}(b(r_0))$ with clipping height $b=b(r_0)$ minimizes
the maximal inefficiency---w.r.t.\ the procedure knowing the respective
radius---among all  procedures ${\rm rLS}(b(r))$, i.e.; each rLS for
some clipping height $b(r)\not=b(r_0)$ has a larger inefficiency for some
$r' \in[r_l,r_u]$.
Radius $r_0$ can be computed quite effectively by a bisection method: Let
\begin{eqnarray}
A_r&=&\Ew_{\rm\SSs id} \Big[ \tr \Cov_{\rm\SSs id}[\Delta X |
                    \Delta  Y^{\rm\SSs id}] +
      (|M^0 \Delta Y^{\rm\SSs id}|-b(r))_+^2\Big] \label{Ardef}\\
B_r&=&\Ew_{\rm\SSs id}\Big[|M^0 \Delta Y^{\rm\SSs id}|^2 -
     (|M^0 \Delta Y^{\rm\SSs id}|-b(r))_+^2\Big] + b(r)^2 \label{Brdef}
\end{eqnarray}
Then $r_0$ solves
  \begin{equation} \label{Lem223b}
 A_{r_0}/A_{r_l}=B_{r_0}/B_{r_u}
  \end{equation}
compare \citet[Lemma~3.1]{Ru:10}.

From another point of view, for given $b$, you may interpret
criteria~\eqref{deltakrit} and \eqref{deltakrit2}
the other way round, giving the efficiency loss in the ideal
model, or the size of the SO neighborhood, for
which the corresponding procedure is MSE-minimax.

\paragraph{(One-Step)-Optimality of the rLS}\label{1stpOpt}
The rLS filter is  optimally-robust in some sense: To see this,
in a first step we boil down our SSM to the following  simplified
model\footnote{Instead of this simplification, we could even
use a more general ``Bayesian'' model as simplification,
compare \citet[section~3.2]{Ru:10}.}. We have an unobservable but
interesting state $X\sim P^X(dx)$, where for technical reasons we assume
that in the ideal model $\Ew |X|^2 <\infty$.
Instead of $X$ we rather observe the sum
\begin{equation} \label{simpAdd}
Y=X+\ve
\end{equation}
of $X$ and a stochastically independent error  $\ve$. We assume
that in the ideal model, the conditional distribution of $Y$
given  $X$ must allow for densities w.r.t.\ some measure $\mu$, i.e.;
\begin{equation} \label{decM}
P^{Y|X=x}(dy)=p^\ve(y-x)\,\mu(dy)\quad \mbox{$P^X(dx)$-a.e.}
\end{equation}
As (wide-sense) AO model,  we consider an SO outlier model already used by
\citet{B:S:93} and \citet{B:P:94}
\begin{equation}
 Y^{\rm\SSs re} = (1-U)  Y^{\rm\SSs id} + U Y^{\rm\SSs di},
 \qquad U\sim {\rm Bin}(1,r) \label{YSO}
\end{equation}
for SO-contamination radius  $0\leq r\leq 1$
specifying the size of the corresponding neighborhood, i.e.; the
probability for an SO. $U$ is assumed independent of
$(X,Y^{\rm\SSs id})$ and $(X,Y^{\rm\SSs di})$ as well as
 \begin{equation}\label{indep2}
 Y^{\rm\SSs di},\; X\quad \mbox{independent}
 \end{equation}
where ${\cal L}(Y^{\rm\SSs di})$  is arbitrary, unknown
and uncontrollable (a.u.u.). The corresponding neighborhood
is defined as
\begin{equation}\label{U-SO}
{\cal U}^{\rm\SSs SO}(r)=\bigcup_{0\leq s\leq r}
\Big\{{\cal L}(X,Y^{\rm\SSs re}) \,|\, Y^{\rm\SSs re} \;
\mbox{acc. to \eqref{YSO} and \eqref{indep2} with radius $s$}\Big\}
\end{equation}
With this setting we may formulate two typical robust optimization problems:
\paragraph{Minimax-SO problem}
Minimize the maximal MSE on an SO-neighborhood, i.e.; find a measurable
reconstruction $f_0$  for $X$ s.t.\
\begin{align}
\quad&\max\nolimits_{{\cal U}}\, \Ew_{\SSs\rm re} |X-f(Y^{\rm\SSs re})|^2 = %
      \min\nolimits_f{}! \label{minmaxSO}
\end{align}
\paragraph{Lemma5-SO problem}
In the spirit of \citet[Lemma~5]{Ha:68}, minimize the MSE in the ideal model
but subject to a bound on the bias to be fulfilled on the whole neighborhood,
i.e.; find a measurable reconstruction $f_0$  for $X$ s.t.\
\begin{align}
\quad& \Ew_{\SSs\rm id} |X-f(Y^{\rm\SSs id})|^2 = \min\nolimits_f{}! \quad
   \mbox{s.t.}\;\sup\nolimits_{\cal U}\big|\Ew_{\SSs\rm re} f(Y^{\rm\SSs re})-
                \Ew X \big|\leq b \label{Lem5SO}
\end{align}
The solution to both problems can be summarized as
\begin{Thm}[Minimax-SO, Lemma5-SO]\label{ThmSO}
\begin{enumerate}
\item[(1)]
In this situation, there is a {\bf saddle-point\/}
     $(f_0, P_0^{Y^{\rm\SSs di}})$ for Problem~\eqref{minmaxSO}
\begin{eqnarray}
f_0(y)&:=&\Ew X +D(y)w_r(D(y)),\qquad w_r(z)=\min\{1, \rho/|z|\} \label{f0def}\\
P_0^{Y^{\rm\SSs di}}(dy)&:=&\Tfrac{1-r}{r} ( \big|D(y)\big|\!/\!\rho\,-1)_{\SSs +}\,\,
P^{Y^{\rm\SSs id}}(dy) \label{P0def}
\end{eqnarray}
where $\rho>0$ ensures that $\int \,P_0^{Y^{\rm\SSs di}}(dy)=1$ and
$D(y)=\Ew_{\SSs\rm id}[X|Y=y]-\Ew X$.
\item[(2)] $f_0$ from \eqref{f0def} also is the solution to
           Problem~\eqref{Lem5SO} for $b=\rho/r$.
\item[(3)] If $\Ew_{\SSs\rm id}[X|Y]$ is linear in $Y$, i.e.;
           $\Ew_{\SSs\rm id}[X|Y]=MY$ for some matrix $M$, then necessarily
\begin{equation}
M=M^0=\Cov(X,Y)\Var Y^{-}
\end{equation}
or in SSM formulation: $M^0$ is just the classical Kalman gain
and $f_0$ the (one-step) rLS.
\end{enumerate}
\end{Thm}
The proof to this theorem is given in \citet[Thm.~3.2]{Ru:10}.

Model~\eqref{simpAdd} already covers our normal SSM model: we only
 have to identify $X$ in model~\eqref{decM} with $\Delta X_t$
and replace $p^\ve(y-x)\,\mu(dy)$ with
 ${\cal N}(Z_t \Delta X_t, V_t)(dy)$. If even $\Delta X_t$ is normal,
 (3) applies and rLS is SO-optimal.

\begin{Rem}\label{rem33}
  \begin{ABC}
\item The  ACM filter by \citet{Ma:Ma:77}, which is a competitor
in this study, by analogy applies \cite{Hu:64}'s minimax variance result
to the ``random location parameter $X$'' setting of \eqref{simpAdd}.
They come up with redescenders as filter $f$. Hence the ACM filter is
not so much vulnerable in the extreme tails but  rather where the
corresponding $\psi$ function takes its maximum in absolute value.
Care has to be taken, as such ``inliers'' producing the least
favorable situation for the ACM are much harder to detect on
na\"ive data inspection, in particular in higher dimensions.
\item For exact SO-optimality of the rLS-filter, linearity of
the ideal conditional expectation is crucial. However, one can
show \citet[Prop.~3.6]{Ru:10} that
$\Ew_{\rm\SSs id}[\Delta  X|\Delta Y]$ is linear iff $\Delta X$
is normal, but, having used the rLS-filter in the $\Delta X$-past,
normality cannot hold, compare \cite[Prop~3.4]{Ru:10}.
 \end{ABC}
\end{Rem}

\paragraph{Back in the $\Delta X$ Model for $t>1$:
           eSO-Neighborhoods}\label{wayoutsec}

As noted in the last remark, rLS fails to be SO-optimal for $t>1$.
Nevertheless rLS performs quite well at both simulations and real data.
One explanation for this is to consider yet an extension of the original
SO-neighborhoods---the {\it e}xtended \textit{SO} or
\textit{${\rm eSO}$--model\/} (compare \citet[Sect.~3.4]{Ru:10}),
where we allow $X$ to be suitably corrupted as well.
In fact, the optimality of pair $(f_0, P_0^{Y^{\rm\SSs di}})$ from
Theorem~\ref{ThmSO} extended
to $\big(f_0, P_0^{Y^{\rm\SSs di}} \otimes P_0^{X^{\rm\SSs di}}\big)$
for any $P_0^{X^{\rm\SSs di}}$ with  $\Ew_{\rm\SSs di} |X^{\rm\SSs di}|^2 = G$,
remains a saddle-point in the corresponding Minimax-Problem on
the ${\rm eSO}$-neighborhood of same radius, cf.\ \cite[Thm.~3.10]{Ru:10}.
As a consequence, (compare~\cite[Prop.~3.11]{Ru:10}), in the
Gaussian setup, instead of focussing on the (SO--) saddle-point
solution to an ${\cal U}(r)$-neighborhood around ${\cal L}(\Delta  X)$
stemming from an rLS-past, we use the following coupling-type idea:

We assume that for each time $t$, there is a
{\em (fictive)} random variable $\Delta  X^{\cal N}\sim{\cal N}_p(0,\Sigma)$
such that $\Delta  X_t^{\rm\SSs rLS}$ stemming from an rLS-past can be
considered a contaminating $X^{\rm\SSs di}$ in the corresponding
${\rm  eSO}$-neighborhood around $\Delta  X^{\cal N}$ with  radius $r$.
From this perspective, the rLS is exactly minimax for each time $t$.

\subsection{IO-optimality} \label{IO'sec}
As noted, in the presence of IO's, we want to follow an IO outlier as
fast as possible. The  Kalman filter in this situation does not behave as
bad as in the AO situation, but still tends to be too inert. To improve upon
this, let us go  back to \eqref{simpAdd} which reveals a useful symmetry
of $X$ and $\ve$: Apparently
\begin{equation} \label{simpEx}
\Ew[X|Y] = Y-\Ew[\ve|Y]
\end{equation}
Hence we follow $Y$ more closely if we damp estimation of $\ve$, for
which we use the rLS-filer. We should note that doing so, we rely
on ``clean'', i.e., ideally distributed errors $\ve$. With the
obvious replacements, Theorem~\ref{ThmSO} translates
word by word to a corresponding minimax Theorem for IO's,
compare \citet[Thm.~4.1]{Ru:10}.

\paragraph{rLS.IO}
In analogy to the definition of the rLS in equation~\eqref{HbDef},
we set up an IO-robust version of the rLS as follows: We retain
the initialization and prediction step of the classical Kalman filter
and, assuming $Z_t$ invertible for the moment, replace the correction
step by
\begin{equation}
X_{t|t} = X_{t|t-1} + Z_t^{-1} [\Delta Y_t -
  H_b\Big((\EM_q-Z_t M^0_t)\Delta Y_t\Big)]
\end{equation}
where the same arguments for the choice of the norm and the clipping
height apply as for the AO-robust version of the rLS.

To better distinguish IO- and AO-robust filters, let us call the
IO-robust version \emph{rLS.IO} and (for distinction) the AO-robust
filter \emph{rLS.AO} in the sequel.

\paragraph{Invertibility problem}
Back in the (linear, discrete-time, Euclidean) SSM the approach
just described faces  the problem that in general matrix $Z_t$ will
not be invertible, so we cannot reconstruct $X$ injectively from
$Y$ and $\ve$. Under a certain full-rank condition, this problem
can be solved by passing to corresponding rLS-type smoothers.
The assumption we need is a version of \emph{complete constructibility},
cf.\  \citet[Appendix]{A:M:79} adopted to the time-inhomogeneous
case:
Denoting the product $F_{t+p}F_{t+p-1}\,\cdot\,\ldots\,\cdot\,F_t$
by $F_{t+p:t}$ we assume that for each $t$, $F_{t+p-1:t}(\R^p)$
is contained in
$[Z_t^\tau, F_{t}^\tau Z_{t+1}^\tau, F_{t+1:t}^\tau Z_{t+2}^\tau,
\ldots,F_{t+p-1:t}^\tau Z_{t+p-1}^\tau](\R^q)$.
Details will be given in a subsequent paper.

\begin{Rem}
\begin{ABC}
\item It is worth noting that also our IO-robust version is a filter,
hence does not use information of observations made after the state
to reconstruct; rLS.IO is strictly recursive and non iterative,
hence well-suited for online applications.
\item An alias to rLS.IO could be \textit{``hysteric filter''\/} as
it hysterically follows any changes in the $Y$'s.
\end{ABC}
\end{Rem}
%
%

\section{Simultaneous Treatment of AO's and IO's} \label{IOAO'sec}
This section is less backed by theoretical results than the ones
on pure AO or IO situation; rather it proposes a heuristic to
achieve both types of outlier robustnesses.
As already mentioned, simultaneous treatment of  AO's and IO's is
only possible with a certain delay. With this delay, we
can base our decision of whether there was an AO or an IO on
the size of subsequent $|\Delta Y_t|$'s---if there was an
AO this should result in only one ``large'' $|\Delta Y_t|$ in a row,
whereas in case of an IO due to propagation, there should be a whole
sequence of large $|\Delta Y_t|$'s. So a hybrid filter
(called rLS.IOAO for simplicity) could be designed as follows:

To a given delay window width $w$, we run in parallel rLS.AO and rLS.IO
(but only store the last $w$ values of rLS.IO). By default we return
the rLS.AO values. Whenever there is a run of $w$ ``large''
$|\Delta Y_t^{\rm\SSs rLS.AO}|$'s we replace the last $w$ filter values
by the corresponding rLS.IO values and use these ones to continue with
the  rLS.AO.

Apparently rLS.IOAO gets into trouble in windows where we have both IO's
and AO's; here some modeling of the type of structural change (local
constant, or better: local linear) as in \citet{F:E:G:07} should be
helpful; we do not treat this in this paper, though.

\paragraph{Operationalization}
In the ideal (Gaussian) model, the $\Delta Y_t$'s should be
independent, so a reasonable decision on whether a sequence of
$|\Delta Y_t^{\rm\SSs rLS.AO}|$'s is ``large'' could be based
on corresponding quantiles of $|\Delta Y_t^{\rm\SSs rLS.AO}|$,
in the ideal model, or, somewhat easier, of
$\Delta Y_t^\tau \Delta_t^{-1} \Delta Y_t$ which, assuming
${\cal L}(\Delta Y_t)$ to be approximately normal, is approximately
$\chi^2_q(0)$-distributed. Relaxing this condition a little,
we already switch to rLS.IO when a high percentage $h$ of the $w$
$|\Delta Y_t|$'s are larger than this given quantile.

This leaves us to determine several tuning parameters:
window-width $w$ (this should really be chosen according to
the application; we have obtained good results in our examples
with $w=5$), the clipping heights for rLS.IO and rLS.AO
(proposal: according to (versions of) \eqref{deltakrit} or
\eqref{deltakrit2}), the percentage $h$ (default: $80\%$ of
the last $w$ instances of $|\Delta Y_t^{\rm\SSs rLS.AO}|$),
and the corresponding quantile (default $99\%$) assuming
that $\Delta Y_t\sim ~ {\cal N}_q(0,\Delta_t)$.

\begin{Rem}
  \begin{ABC}
  \item Note that although the decision whether we issue the rLS.IO or
    the rLS.AO values is made $w$ observations after the state which
    is to be reconstructed, we still only use filters, hence the
    information of $Y_{t+j}$, $j=1,\ldots,w-1$ is not used to
    improve the reconstruction so far, as this would involve corresponding
    (yet-to-be-robustified) smoothers.
    Once the corresponding work on robust smoothing will be done, we
    could surely use this additional information.
  \item In a time-invariant linear SSM (i.e., with hyperparameters
    $F$, $Z$, $Q$, and $V$ constant in time), in general $\Delta_t$ will
    converge in $t$ exponentially fast---also if one uses bounded
    rLS-steps---so these tuning parameters will only have to be
    determined for a small number of time instances $t$, compare \citet[chap.~7]{Ru:01}.
    In fact, setting them time-invariant right from the beginning
    will often do a reasonable job already.
  \end{ABC}
\end{Rem}
%
\section{Simulation Example: Steady State Model}\vspace{-.5ex} \label{SimulStud}
%
Our running example will be a one-dimensional steady state model with
hyper-parameters
\begin{equation}\label{steadystate}
p=q=1,\quad F_t=Z_t=1, \qquad\mbox{in the ideal model: } v_t, \ve_t \iid {\cal N}(0,1)
\end{equation}
We consider performance of classical Kalman filter, rLS.AO, rLS.IO,
and rLS.IOAO  in this model and under AO's and IO's.
More specifically, we have generated deterministic
AO's in observations 10,15,23, and IO's in observations
20--25 (a local linear trend) and 37--42 (level shift).

As competitors, we include the ACM filter by \citet{Ma:79}
implemented by B.~Spangl in {\sf R} package {\tt robKalman}, and a variant
$\mbox{hybr}_{\SSs\mbox{\scriptsize PRMH}}$ of {\tt robfilter}, cf.\
\citet{F:S:08} concerning its implementation and \citet{F:B:G:06}
regarding its definition, which is a non-parametric filter fitting local
levels and linear trends based on repeated medians.
The results are plotted in Figures~\ref{Fig3}--\ref{Fig6},
where in the plots, we confine ourselves to the rLS-variants,
which already makes for five curves to be plotted in one panel.

In the ideal situation, all filters perform well, with
slight advantages for the classical Kalman filter (which has
smallest theoretical MSE), but closely followed
(and in the prediction case slightly beaten) by the rLS.IO.

In the IO situation,
 the ``hysteric'' rLS.IO filter performs best, beating
 the classical Kalman filter; both rLS.IOAO and
 $\mbox{hybr}_{\SSs\mbox{\scriptsize PRMH}}$ perform reasonably well,
 while the AO-robust filters ACM and rLS.AO are not able to track
 the IO at all (as they can only perform bounded correction steps)
 and hence, like a hanging slope, only closely recover the
 changed situation.

 In the AO situation, we have the complementary image;
 here ACM performs best (see also Remark~\ref{rem61}(a)), but
 rLS.AO only performs slightly weaker. rLS.IOAO is a little worse,
 and with a certain gap, but still reasonably well follows
 $\mbox{hybr}_{\SSs\mbox{\scriptsize PRMH}}$, while both
 classical Kalman filter and rLS.IO
 (the latter even worse) perform drastically bad.

 Finally, in the mixed IO and AO situation,
 $\mbox{hybr}_{\SSs\mbox{\scriptsize PRMH}}$ is by far the
 best solution, then followed with a certain gap by the
 rLS.IOAO, while all other filters perform unacceptably bad.
 By construction, rLS.IOAO assumes that at every time instance
 there only can be either an AO or an IO.
 Otherwise the corresponding MSE gets unbounded on every
 neighborhood ${\cal U}(r)$ for $r>0$. Hence the AO in observation
 23 really confuses rLS.IOAO completely, compare
 Figures~\ref{Fig3}--\ref{Fig6}:  it has just switched to
 ``hysteric'' IO behavior and hence faithfully follows the AO.
 Repeated-median-based $\mbox{hybr}_{\SSs\mbox{\scriptsize PRMH}}$
 does not have this problem, as the median even stays stable under
 (almost) arbitrary substitutive outliers, hence it is able to
  keep the local linear trend. Omitting observation 23 results
  in a much better performance of rLS.IOAO, which then even
  beats $\mbox{hybr}_{\SSs\mbox{\scriptsize PRMH}}$, cf.\
  Table~\ref{tab2}.

 \begin{figure}
\centerline{\includegraphics[height=9cm, width=12cm]{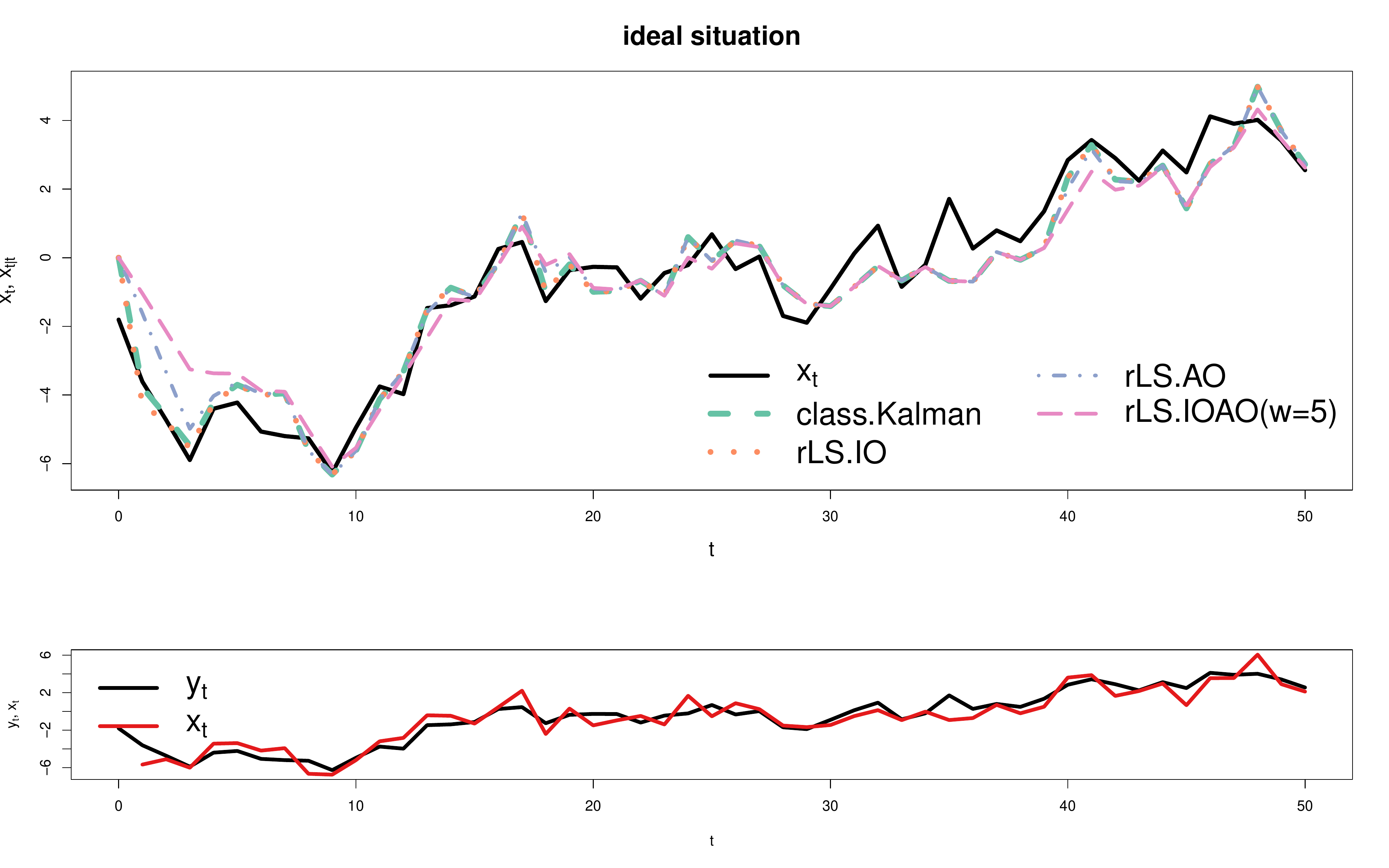}}
\caption{\label{Fig3}rLS-filter variants in model~\eqref{steadystate}
in the ideal model; in the panel below (note the different $y$-scale)
both actual states (black) and observations (red) are plotted.}
\end{figure}

\begin{figure}
\centerline{\includegraphics[height=9cm, width=12cm]{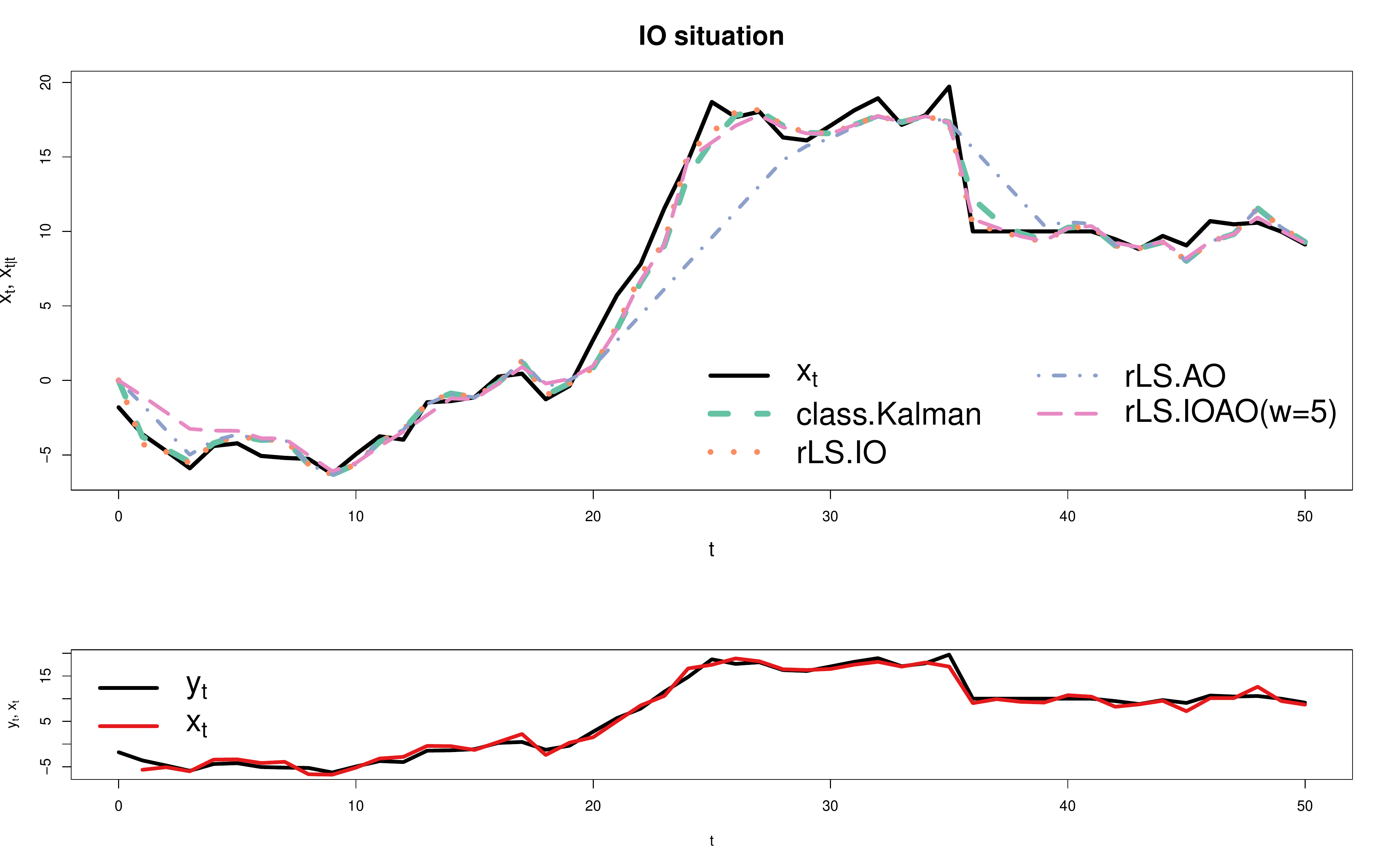}}
\caption{\label{Fig4}rLS-filter variants in model~\eqref{steadystate}
with IO's: a local linear trend at $X_{20}$--$X_{25}$ and a level shift
for states $X_{37}$--$X_{42}$; the panel below (note the
different $y$-scale) is as in Figure~\ref{Fig3}.}
\end{figure}

\begin{figure}
\centerline{\includegraphics[height=9cm, width=12cm]{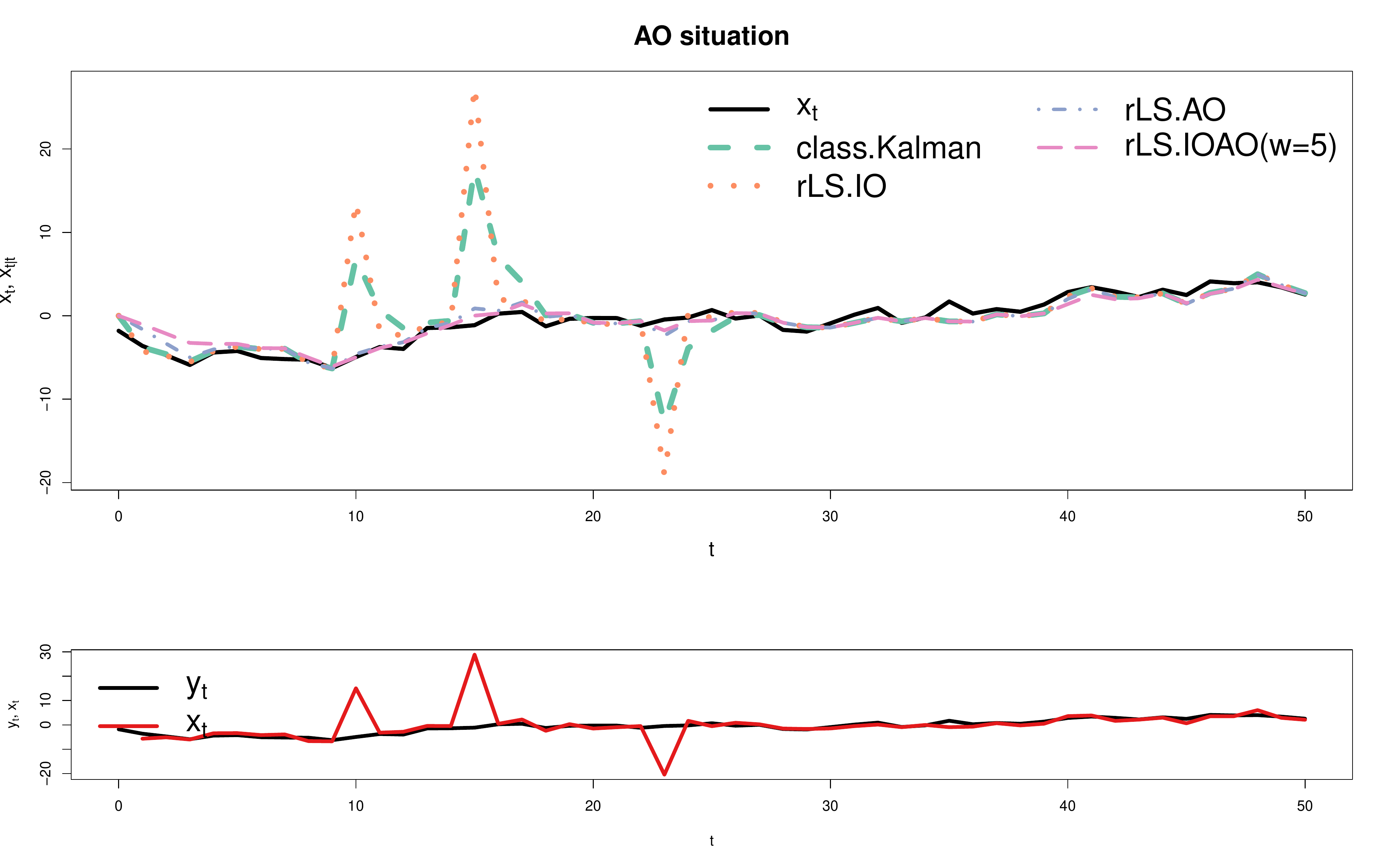}}
\caption{\label{Fig5}rLS-filter variants in model~\eqref{steadystate}
with AO's in observations 10,15,23; the panel below (note the
different $y$-scale) is as in Figure~\ref{Fig3}.}
\end{figure}

\begin{figure}
\centerline{\includegraphics[height=9cm, width=12cm]{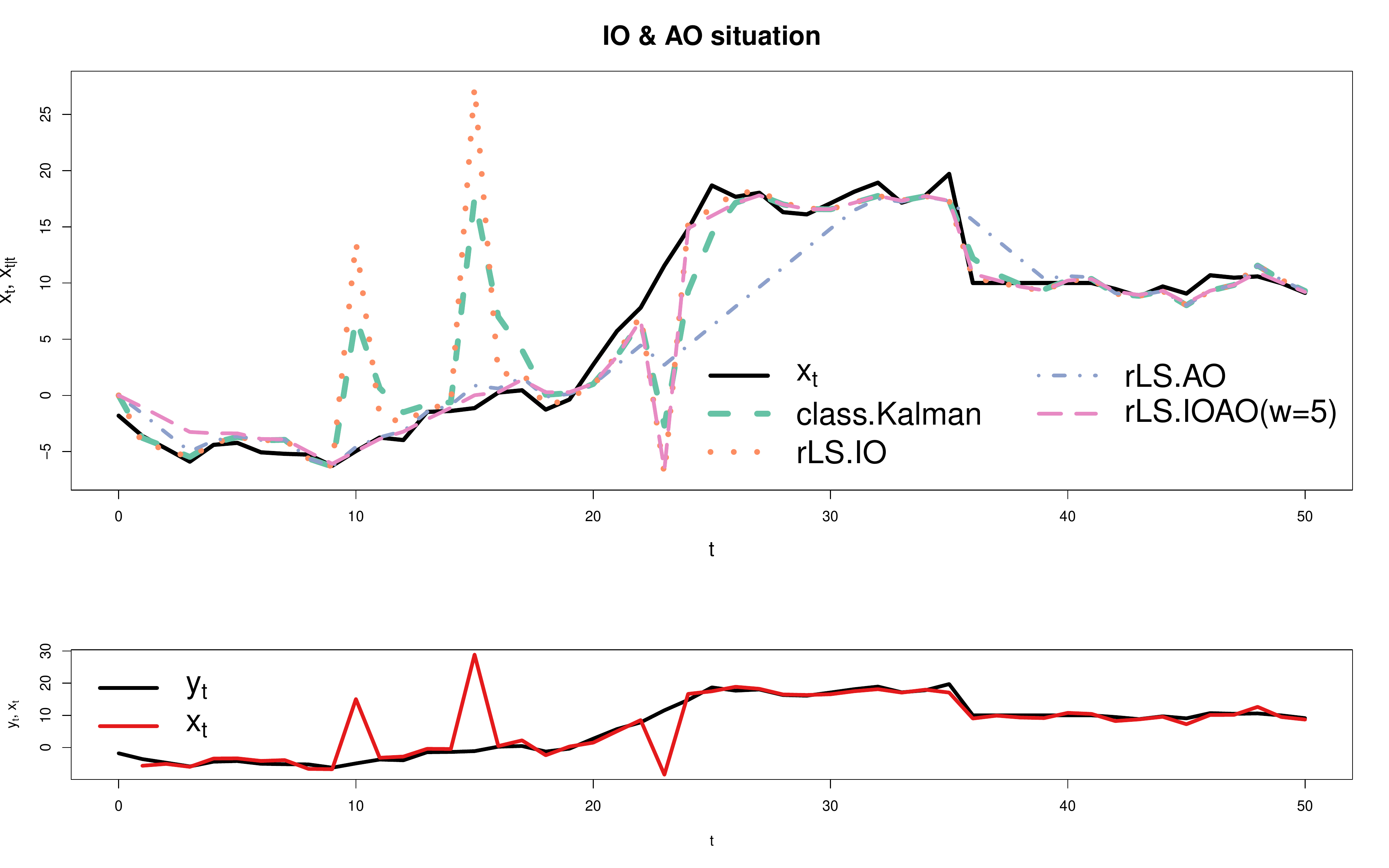}}
\caption{\label{Fig6}rLS-filter variants in model~\eqref{steadystate}
with both  IO's as in Figure~\ref{Fig4} and AO's as in Figure~\ref{Fig5};
the panel below (note the different $y$-scale) is as in Figure~\ref{Fig3}.}
\end{figure}

Averaging over time in one realization of the SSM, we get the ``ergodic''
empirical MSEs as displayed in Tables~\ref{tab1},~\ref{tab2}.

\begin{table}
\begin{center}
\begin{tabular}{ll|rrrr|rr}
\multicolumn{8}{c}{\rule[-.4ex]{0mm}{3ex}{empirical MSE}}\\[2ex]
Situation&Type&Kalman&$\mbox{rLS}_{\mbox{\scriptsize IO}}$&%
$\mbox{rLS}_{\mbox{\scriptsize AO}}$&$\mbox{rLS}_{\mbox{\scriptsize IOAO}}$
&$\mbox{ACM}$&$\mbox{hybr}_{\SSs\mbox{\scriptsize PRMH}}$\\
\hline
\multirow{2}*{ideal}   & filter&     \textbf{ 0.59}&             0.60 &   0.75&            1.08 &          0.77 &           1.41 \\
                            &pred&            1.69 &    \textbf{ 1.67}&   1.96&            2.26 &          2.01 &                \\
\hline\multirow{2}*{IO}      & filter&        1.04 &    \textbf{ 0.83}&   6.54&            1.36 &         25.19 &           1.36 \\
                            &pred&            5.28 &    \textbf{ 4.71}&  12.17&            5.42 &         32.16 &                \\
\hline\multirow{2}*{AO}      & filter&       15.25 &            30.38 &   0.91&            1.16 &  \textbf{0.82}&           1.79 \\
                            &pred&           15.15 &            29.68 &   2.00&            2.25 &  \textbf{2.05}&                \\
\hline\multirow{2}*{IO\&AO }& filter&        17.00 &            30.52 &  12.89&   \textbf{ 7.78}&         28.76 &   \textbf{1.53}\\
                            &pred&           21.94 &            34.56 &  19.23&   \textbf{13.87}&         36.08 &                \\
\end{tabular}
\caption{\label{tab1}``ergodic'' estimates for the MSE of the variants of the
rLS and the ACM and $\mbox{hybr}_{\SSs\mbox{\scriptsize PRMH}}$
in the situation described in the text; best results are printed in bold face.}
\end{center}
\end{table}

\begin{Rem}\label{rem61}
\begin{ABC}
  \item We can explain why the ACM filter beats the rLS in the AO-situation
    by the fact that the contamination in this study clearly covers the
    worst-case behavior of the rLS but not the one of the ACM filter,
    compare Remark~\ref{rem33}(e), and also fails to do so for
    $\mbox{hybr}_{\SSs\mbox{\scriptsize PRMH}}$.
  \item rLS.IOAO really has its advantages in higher dimensions where
    median-based filters are much harder to define and get computationally
    expensive. One might even think of combining rLS.IOAO and
    $\mbox{hybr}_{\SSs\mbox{\scriptsize PRMH}}$ in these settings:
    first let rLS.IOAO do a preliminary, fast, and dimension-independent
    cleaning, and then let $\mbox{hybr}_{\SSs\mbox{\scriptsize PRMH}}$
    polish this result coordinate-wise.
  \item It is still an open question whether we can improve on
    the rLS.IOAO behavior, using the SSM
  \begin{equation}\nonumber
  Z_t=(1,t), X_t=(a_t,b_t)^\tau, F_t=\EM_2, Q_t=0.1\EM_2, V_t=1
  \end{equation}
    which (up to the specification of error/innovation variance) is
    essentially the model in the background of
    $\mbox{hybr}_{\SSs\mbox{\scriptsize PRMH}}$.
    In this setting $Z_t$ is not invertible, but the model is
    completely constructible, so passing to smoothers might help.
\end{ABC}
\end{Rem}

\begin{table}
\begin{center}
\begin{tabular}{ll|rrrr|rr}
\multicolumn{8}{c}{\rule[-.4ex]{0mm}{3ex}%
  {{empirical MSE}}---without obs.\ $23$}\\[2ex]
Situation&Type&Kalm&$\mbox{rLS}_{\mbox{\scriptsize IO}}$&%
  $\mbox{rLS}_{\mbox{\scriptsize AO}}$&$\mbox{rLS}_{\mbox{\scriptsize IOAO}}$
  &$\mbox{ACM}$&$\mbox{hybr}_{\SSs\mbox{\scriptsize PRMH}}$\\
\hline
\multirow{2}*{ideal}   & filter&     \textbf{ 0.59}&              0.60 &   0.75&            1.10 &          0.78 &           1.43 \\
                            &pred&            1.71 &    \textbf{  1.69}&   1.99&            2.29 &          2.03 &                \\
\hline\multirow{2}*{IO}      & filter&        0.94 &    \textbf{  0.74}&   6.08&            1.26 &         24.48 &           1.38 \\
                            &pred&            5.59 &    \textbf{  4.98}&  12.05&            5.73 &         31.66 &                \\
\hline\multirow{2}*{AO}      & filter&       12.46 &             24.07 &   0.86&            1.15 & \textbf{ 0.84}&           1.83 \\
                            &pred&           12.18 &             23.05 &   1.94&            2.25 & \textbf{ 2.10}&                \\
\hline\multirow{2}*{IO\&AO }& filter&        13.28 &             24.21 &  11.58&   \textbf{ 1.31}&         27.93 &           1.56 \\
                            &pred&           17.01 &             26.34 &  17.80&   \textbf{ 5.63}&         35.38 &                \\
\end{tabular}
\caption{\label{tab2} results as in Table~\ref{tab1}, but excluding the
values for observation 23, where we had coincidence of (wide-sense)
IO and AO, a situation not covered in the design of rLS.IOAO.}
\end{center}
\end{table}
\begin{Rem}
A careful reader might object that we have not included any real world data application.
We have done so, because most data sets we have analyzed come with
unkown hyper-parameters. The setting of this paper assumes knowledge
of these hyper-parameters, so we would have to estimate them somehow---possibly by
the EM algorithm of \citet{Sh:St:82} or again by more recent refinements.
But, since the robustness properties of combining a non-robust M-step
(i.e., maximum likelihood) and a robust E-step (achieved using
the procedures of this paper) are not evident, and a corresponding
robustification of the M-step would have been out of scope,
we have rather confined us to simulations.
\end{Rem}
\section[Implementation: R-package robKalman]%
   {Implementation: {\sf R}-package {\tt robKalman}} \label{implem}
rLS.AO was originally implemented to {\tt XploRe}, compare \citet{Ru:00}.
In an ongoing project with Bernhard Spangl, BOKU, Vienna,
and Irina~Ursachi (ITWM),
we are about to implement all the rLS filter to {\sf R}, see \citet{R09},
more specifically to an {\sf R}-package {\tt robKalman}, the development
of which is done under {\tt r-forge} project
\hreft{https://r-forge.r-project.org/projects/robkalman/},
see also \citet{Rforge}. Under this address you will also find a
preliminary version available for download.

\section{Conclusion} \label{conclusion}
In the extremely flexible class of dynamic models consisting in SSMs, we
apply (distributional) robust optimality results for filtering.
We could show that contrary to common prejudice  a simultaneous treatment
of (wide-sense) IO's and AO's is possible in SSM's---albeit with minor
delay. 
 The filters that we propose are model based (in contrast to
the non-parametric $\mbox{hybr}_{\SSs\mbox{\scriptsize PRMH}}$) which
means that we need a higher degree of model specification in that we
possibly have to estimate the hyper-parameters, but which also could
help to get more precise in the ideal model (in particular for higher
dimensions). 
Our filters are non-iterative, recursive, hence fast, and valid for
higher dimensions. 
They are available in {\sf R} in some devel versions and
hopefully on {\tt CRAN} soon.
\section*{Acknowledgements}
The author would like to acknowledge and thank for the stimulating
discussion he had with Gerald Kroisandt at ITWM which led to the
definition of rLS.IO. Many thanks go to
Nataliya Horbenko for proof-reading the manuscript.
\section*{References}
\bibliographystyle{plainnat}

\end{document}